\documentclass[multphys,vecphys]{svmult}

\usepackage{makeidx}         
\usepackage{graphicx}        
\usepackage{multicol}        
\usepackage[bottom]{footmisc}

\makeindex             

\begin{document}

\title*{Integral Field Spectroscopy of the core of Abell 2218}
\author{S.F.S\'anchez\inst{1}\and
N.Cardiel\inst{1,2}\and
M.Verheijen\inst{3}\and
N.Benitez\inst{4}
}
\institute{Centro Astron\'omico Hispano Alem\'an, Calar Alto, (CSIC-MPI),
C/Jesus Durban Remon 2-2, 04004-Almeria, Spain,
\texttt{sanchez@caha.es}
\and Departamento de Astrof\'\i sica, Facultad de F\'\i sicas, Universidad Complutense de Madrid, 28040 Madrid, Spain \texttt{cardiel@caha.es}
\and Kapteyn Astronomical Institute, PO Box 800, 9700 AV Groningen, the Netherlands
\and Instituto de Astrof\'\i sica de Andalucia CSIC, Camino Bajo de Huetor
S/N, Granada, Spain
}
%
%
\maketitle

\begin{abstract}
We report on integral field spectrocopy observations, performed with
the PPAK module of the PMAS spectrograph, covering a field-of-view of
$\sim$74''$\times$64'' centered on the core of the galaxy cluster
Abell 2218. A total of 43 objects were detected, 27 of them galaxies
at the redshift of the cluster. We deblended and extracted the
integrated spectra of each of the objects in the field using an
adapted version of {\tt galfit} for 3D spectroscopy ({\tt
galfit3d}). We use these spectra, in combination with morphological
parameters derived from deep HST/ACS images, to study the stellar
population and evolution of galaxies in the core of this cluster.
\end{abstract}

\section{Introduction}
\label{sec:1}

Galaxy clusters have been used for decades to study the evolution of galaxies.
Being tracers of the largest density enhancements in the universe, clusters
are considered the locations where galaxies formed first. It is known that
they are dominated by old and large elliptical galaxies, with colors that are
consistent with a bulk formation at high redshift followed by a passive
evolution (e.g., \cite{zieg01}). However, \cite{butc84} have shown an increase
in the fraction of blue galaxies in clusters from low to intermediate
redshift, which disagrees with that simple scenario. Furthermore, a fast
morphological evolution from late to early-type galaxies, claimed as a
possible solution, does not predict the observed fractions of S0 galaxies at
low redshift. Other processes like gas-stripping and gravitational harassment
have to be considered. These notions predict that galaxies at the core of
clusters must have globally older stellar populations than galaxies in the
outskirts of the cluster.  Consequently, we should detect deviations from
passive evolution in the scaling relations for early-type galaxies (e.g., the
Fundamental Plane). In order to test this hypothesis we started a complete
spectroscopic survey of the core of Abell 2218.

\begin{figure}
\centering
\includegraphics[height=5.5cm]{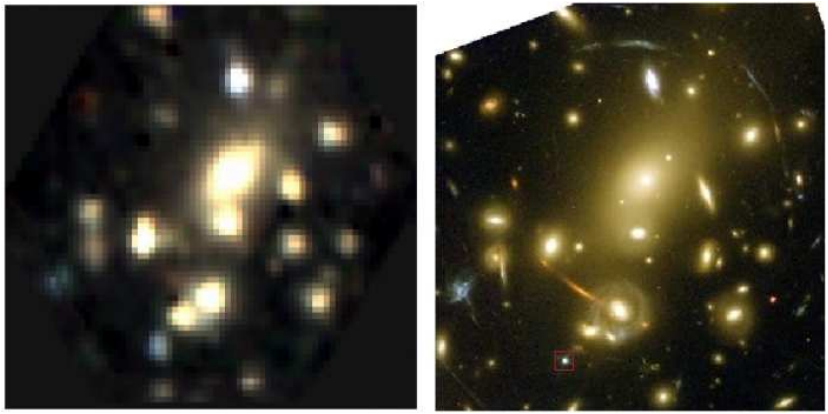}
%
%
\caption{{\it left-panel:} Three-color image created by coadding the
flux of the final datacube through three broad-bands corresponding
approximately to $V$,$R$ and $I$. {\it right-panel:} Similar image
created using HST data (Fruchter et al.,\cite{acs}).
The star used to derive the PSF is indicated with a red square.}
\label{fig:1}       
\end{figure}

Abell 2218 is one of the richest clusters in the Abell catalogue
\cite{abell89}, with a richness class 4. It has a redshift of
$z\sim$0.17, and a velocity dispersion of 1370 km s$^{-1}$
\cite{kris78,borg92}. Detailed high-resolution X-ray maps
\cite{mcha90} and mass-concentration studies based on the properties
of the gravitationally lensed arcs have shown that the cluster
contains two density peaks, the strongest of them dominated by a large
cD galaxy. Several observational programs have produced a large
dataset of well deblended slit spectra for the galaxies in the outer
parts of this cluster \cite{zieg01}, and extensive multi-band,
ground-based and HST imaging, e.g., \cite{acs}. We focused our survey
in the central arcmin region of the cluster, around the cD galaxy.

\subsection{Observations and data reduction}
\label{sec:2}

Observations were carried out on 30/06/05 and 06/07/05 at the 3.5m
telescope of the Calar Alto observatory with the PMAS \cite{roth05}
spectrograph and its PPAK module \cite{kelz06}. The V300 grating was
used, covering a wavelength range between 4687-8060 \AA\ with a
nominal resolution of $\sim$10 \AA\ FWHM. The PPAK fiber bundle
consists of 331 science fibers of 2.7'' diameter, concentrated in a
single hexagonal bundle covering a field-of-view of
72''$\times$64''. Following a dithered 3-pointing scheme, 3 hours of
integration time was accumulated each night, 6 hours in total. In
addition to our Integral Field Spectroscopy (IFS) data, we used a
F850LP-band image of 11310s exposure time taken with the ACS camera on
board the HST, obtained from the HST archive.

Data reduction was performed using R3D\cite{sanc05}, in combination
with IRAF packages and E3D\cite{sanc04}. The reduction involved
standard steps for fiber-based integral-field spectroscopy. First,
science frames are bias subtracted. A continuum illuminated exposure,
taken before the science exposures, is used to locate and trace the
spectra on the CCD. Each spectrum is then extracted by coadding the
flux within an aperture of 5 pixels.
Wavelength and flux calibration are performed using an arc lamp and standard
calibration star exposures, respectively. The three dithered exposures are
then combined, and a datacube with 1''/pixel sampling is created for each
night using E3D. Finally, the two datacubes are recentered and combined using
IRAF tasks.

\begin{figure}
\centering
\includegraphics[height=4.5cm]{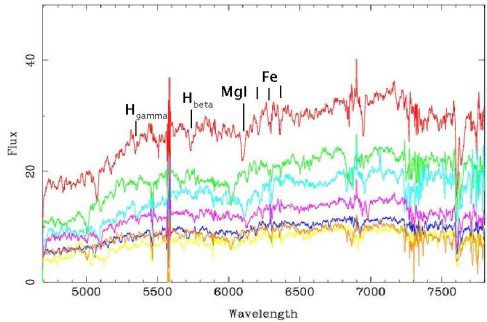}
%
%
\caption{Example of the dataset,
  showing the spectra of some of the brightest galaxies
  in the cluster.
  Some absorption features of interest are indicated.
}
\label{fig:3}       
\end{figure}

Figure \ref{fig:1} shows a three-color image created by coadding the flux of
the final datacube through three broad-bands: $V$,$R$ and $I$-band (left
panel), together with a similar image created using HST observations (right
panel)\cite{acs}. It is interesting to note the similarities between the two
images, despite the differences in sampling and resolution.  As we quoted
before, many galaxies are strongly blended. In particular, the central cD
contaminates most of the galaxies in the field.

\subsection{Galaxy detection and spectra extraction}
\label{sec:3}

In order to deblend and extract the integrated spectra of each
individual galaxy in the field, we used a technique developed by
ourselves \cite{sanc04a,garc05}. The technique is an extension to IFS
of {\tt galfit} \cite{peng02}, that we named {\tt galfit3d}. It
entails a deblending of the spectrum of each object in the datacube by
fitting analytical models. IFS data can be understood as a set of
adjacents narrow-band images, each with the width of a spectral
pixel. For each narrow-band image it is possible to apply modelling
techniques developed for 2D imaging, like {\tt galfit}, and extract
the morphological and flux information for each object in the field at
each wavelength. The spectra of all the objects are extracted after
repeating the procedure for each narrow-band image throughout the
datacube.  We have already shown that the use of additional information to
constrain the morphological parameters increases the quality of the
recovered spectra \cite{garc05}. For that purpose we have used the
F850LP-band image obtained with the HST/ACS camera.

First, we use SExtractor \cite{bert96} on the section of the
F850LP-band image correponding to the field-of-view of our IFS
data. For each detected galaxy we recover its position, integrated
magnitude, scale length, position angle and ellipticity.  These
parameters were used as an initial guess to fit each of the galaxies
in the F850LP-band image with a 2D S\'ersic profile model, convolved
with a PSF, using {\tt galfit}. The PSF was obtained using a poststamp
image of the star in the PPak field-of-view. The fit for each individual
object was done in a sequential way, from the brightest to the
faintest, masking all the remaining objects. After iterating over all
the detected galaxies we obtain a final catalogue of their
morphological parameters. Similar techniques are used to derive the
morphological parameters of galaxies in different ACS imaging surveys
\cite{rix04,coe06}.

\begin{figure}
\centering
\includegraphics[height=6cm]{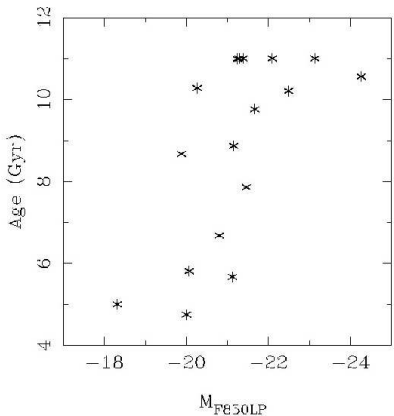}
%
%
\caption{Luminosity-weight age of some galaxies in the cluster core
  along the absolute magnitude in the F850LP-band.  }
\label{fig:4}       
\end{figure}

The catalogue of detected galaxies in the F850LP-band image was
cross-checked visually with a 2D image derived from the datacube by
coadding the flux over the entire wavelength range (4687-8060 \AA).  A
final catalogue of 41 objects (40 galaxies and 1 star) was created,
excluding two detected arc-lenses. Once we derived the morphological
parameters for each galaxy, we extracted their integrated spectra by
modelling each galaxy in the datacube with {\tt galfit3d}. We use the
same model used for the F850LP-band image, with the morphological
parameters fixed, and fitting only the flux at each wavelength. The
fit was performed again in a sequential way, from the brightest to the
faintest, masking all the objects but the fitted one. In each
iteration we used as input the residual datacube of the previous
iteration, ``cleaning'' each object one-by-one. We finally got a
spectrum for each of the 40 galaxies in the field-of-view.

\subsection{Analysis and Results}
\label{sec:4}

Figure \ref{fig:3}, shows a few examples of the extracted
spectra. Different absorption features are detected in each spectrum,
including some that are sensitive to age and metallicity (e.g.,
H$\beta$ and MgI). A few spectra show clear gaseous emission lines
(e.g., H$\beta$, [OIII] and H$\alpha$). We derive redshifts for 28 of
the 40 galaxies by comparing the observed wavelength of the absorption
(or emission) features with the restframe values. Of those, 27
galaxies are at a redshift around $z\sim0.17$, the nominal redshift of
the cluster, and another one at $z=0.104$. The signal-to-noise of the
remaining 12 spectra is too low to unambiguously identify spectral
features. We derived the age and metallicity of the galaxies by
fitting each spectrum with single stellar population synthetic models,
created using the GISSEL code \cite{bc04}.

Figure \ref{fig:4} shows a preliminary result obtained from the
combination of parameters derived from the fitting procedure and the
morphological analysis. It shows the luminosity-weighted age versus
the F850LP-band absolute magnitude for the 17 galaxies for which we
have completed the analysis. There is a large spread of ages in the
stellar populations of the galaxies in the core of Abell 2218,
contrary to the expectations from a single bulk formation and a
passive evolution. Furthermore, the brightest (and more massive)
galaxies are older than the fainter (and less massive) ones, which
show a larger spread in ages. This may indicate that the smaller, less
massive galaxies have formed later, being captured by the cluster,
and/or they have enjoyed more recent periods of star formation. These
results illustrate that the evolution of stellar populations in
galaxies in clusters is far from passive, even in the central core.



\printindex
\end{document}